\documentclass[useAMS,usenatbib]{mn2e}
\usepackage{graphicx}
\usepackage[hyperindex,breaklinks]{hyperref}
\def\jcoph{J. Comp.\ Phys.}

\def\eg{{\it e.g.}}
\def\etal{{\it et al.}}
\def\etc{{\it etc.}}
\def\ie{{\it i.e.}}

\def\DF{{\small DF}}

\def\pmb#1{\setbox0=\hbox{$#1$}%
  \kern-0.25em\copy0\kern-\wd0
  \kern.05em\copy0\kern-\wd0
  \kern-0.025em\raise.0433em\box0}

\def\ILR{{inner Lindblad resonance}}
\def\OLR{{outer Lindblad resonance}}

\title[Short-lived Spirals in Disc Galaxies]{The Lifetimes of Spiral
  Patterns in Disc Galaxies}
\author[J. A. Sellwood]{J. A. Sellwood$^{1}$\thanks{E-mail:
    sellwood@physics.rutgers.edu} \\ $^{1}$Rutgers University,
  Department of Physics \& Astronomy, 136 Frelinghuysen Road,
  Piscataway, NJ 08854-8019, USA}

\begin{document}


\pagerange{\pageref{firstpage}--\pageref{lastpage}} \pubyear{2010}

\maketitle

\label{firstpage}

\begin{abstract}
The rate of internally-driven evolution of galaxy discs is strongly
affected by the lifetimes of the spiral patterns they support.
Evolution is much faster if the spiral patterns are recurrent
short-lived transients rather than long-lived, quasi-steady features.
As rival theories are still advocated based on these two distinct
hypotheses, I review the evidence that bears on the question of the
lifetimes of spiral patterns in galaxies.  Observational evidence from
external galaxies is frustratingly inconclusive, but the velocity
distribution in the solar neighbourhood is more consistent with the
transient picture.  I present simulations of galaxy models that have
been proposed to support quasi-steady, two-arm spiral modes that in
fact evolve quickly due to multi-arm instabilities.  I also show that
all simulations to date manifest short-lived patterns, despite claims
to the contrary.  Thus the transient hypothesis is favoured by both
numerical results and the velocity distribution in the solar
neighbourhood.
\end{abstract}

\begin{keywords}
galaxies: evolution -- galaxies: kinematics and dynamics --
galaxies: spiral
\end{keywords}

\section{Introduction}
After bars, spirals are the most prominent features of galaxy discs.
The question of whether the spirals are short- or long-lived features
is of major importance to the internal evolution of disc galaxies.  In
this paper, I evaluate the evidence that spirals are transient features
that change within a few dynamical times, but are superseded by fresh
patterns in a recurrent manner.

We still lack a complete, and widely accepted, theory for the origin
of spiral patterns in galaxies.  There is good evidence
\citep[\eg][]{KN79} that many prominent spiral patterns are found in
barred galaxies, or are driven by tidal interactions, or perhaps even
DM halo substructure \citep[\eg][]{Dubi08}.  There seems little doubt
that a tidally-driven spiral pattern evolves rapidly
\citep[\eg][]{SL93,Dobb10}, but repeated excitations may be possible
\citep{BH92}.  If bars are long-lived features, as is generally
believed \citep[for a dissenting view see][]{BCS05}, spirals driven by
bars could also be long-lived patterns.  However, simulations
\citep{SS88} suggest that spirals and bars could have differing
pattern speeds, which seemed to be supported by observational data
\citep{SCJ03,Buta09}, although the latter group recently changed their
minds \citep{Salo10}!

Although many spirals in galaxies could be driven responses, the
ubiquity of the spiral phenomenon suggests others are likely to be
self-excited features of discs.  In particular, all driving agents can
be excluded from $N$-body simulations of isolated stellar discs, which
continue to manifest spiral patterns that must be self-excited.  A
satisfactory theory to account for self-excited spirals has yet to
emerge, despite decades of effort \citep[see reviews
  by][]{Toom77,Atha84,BL96,Sell10a}.  While all agree spirals are
gravitationally-driven variations in the surface density of the old
stellar disc, there is no consensus even on the expected lifetimes of
the patterns.

Here I first explain why the duration of spiral features is an
important issue, and then elaborate on the brief discussion of the
lifetimes of spirals presented by \citet[][hereafter BT08,
  p.~526]{BT08}.  I evaluate four different types of evidence that
bear on the question whether spirals are short- or long-lived
patterns.

\section{Internal Evolution of Galaxy Discs}
\label{evolv}
It is not obvious that the rate of evolution of galaxy discs is
affected by the lifetimes of the patterns.  It may seem that if the
time-averaged amplitude and pitch angle of transient spirals does not
differ much from those of a steady long-lived pattern, the rate of
change in the distribution of angular momentum and other quantities
would be similar.

However, the rate of angular momentum transport by spiral waves is
given by the rate at which wave action \citep{LBK72} is transported at
the group velocity \citep{Toom69}, which can differ from the gravity
torque by the advective Reynolds stress (BT08, Appendix J), also known
as ``lorry transport.''  Thus angular momentum transport is
substantially reduced for long-lived patterns that invoke feed-back
via the long-wave branch of the dispersion relation \citep{Mark77},
where the advective term actually reverses the sign of angular
momentum transport \citep{Sell10a}.  Transient spirals do not
involve the long-wave branch, and therefore redistribute angular
momentum more effectively.

Also, \cite{LBK72} showed that stars are scattered by a slowly
changing potential perturbation only near resonances.  More precisely,
a spiral potential that grows and decays adiabatically, \ie\ on a
time-scale long compared with the orbital and epicyclic periods, will
not cause a lasting change to a star's orbit.  Lasting changes do
arise where wave-particle interactions become important near the
resonances.  Stars experience secular changes through ``surfing'' on
the potential variations at corotation or through a periodic forcing
close to their epicyclic frequency at the Lindblad resonances.

Since exchanges between the stars and the wave take place at
resonances, large changes induced by a long-lived pattern would be
confined to a few narrow resonances, the most important of which, the
inner Lindblad resonance, must be ``shielded'' to avoid fierce damping
of the spiral \citep{Mark74,BL96}.  The frequency widths of resonances
of a steady wave depend only on the amplitude of the perturbation.
But resonances are broader for time-dependent waves, and multiple,
short-lived disturbances having a range of pattern speeds change the
angular momenta of stars over large parts of the disc.

Any mechanism to excite spirals relies on the extraction of angular
momentum from stars in the inner disc and its deposition on stars in
the outer disc \citep{LBK72}.  Resonant stars of a finite-amplitude
wave will change their orbits as a result of these angular momentum
exchanges and a spiral wave must decay once the resonances, which are
the sources and sinks that excite it, have been exhausted.
\cite{Bert83} found order unity angular momentum changes would take a
few Hubble times for a spiral of 20\% overdensity with a pitch-angle
$\sim 16$\degr.  But resonances will be depopulated on a much shorter
time-scale for large-amplitude, open spirals, which therefore cannot
last long if the pattern speed is constant.  Note that the lifetime
could be lengthened by a changing pattern speed, which would allow the
resonances to sweep over the disc.

Thus the question of the lifetimes of spirals is of major importance
to our understanding of disc evolution.  If, on the one hand, the
recurrent transient picture is correct, spirals are the most important
agents driving the evolution of galaxy discs: they transport angular
momentum, scatter stars into non-circular orbits, and cause radial
mixing (see also \S\ref{indirect}).  Long-lived, quasi-steady
patterns, on the other hand, which must have moderate amplitude and be
tightly wrapped in order to persist, would have a much less
significant effect.

\section{Observational Evidence}
\label{observ}
Near-IR photometric images
\citep{Schw76,Bloc94,GGF95,RZ95,SJ98,Eskr02,GPP04} reveal that spiral
patterns represent large amplitude variations in the surface
brightness of the old stellar disc.  \cite{ZCR09} attempt to derive a
more faithful mass map from their H-band images by applying a pixel by
pixel correction based on population synthesis models for the
broad-band colours.  All this work has confirmed that spiral patterns
constitute large-amplitude density variations in the total stellar
density in the disc, but says nothing about the lifetimes of the
features.

Velocity maps of galaxies that can resolve inter-arm variations reveal
that spiral patterns are associated with organized non-axisymmetric
streaming flows \citep[\eg][]{Viss78,CAT93,Shet07}.  The amplitude of
the non-axisymmetric potential variations can be estimated by modeling
such data \citep[\eg][]{KSR03}, confirming that spirals are associated
with significant potential variations.  But unfortunately, these
measurements are again quite insensitive to the lifetimes of the
spirals; the flow pattern is essentially the gas response to the
instantaneous potential variation, and its qualitative features are
independent of the mechanism that created it (BT08, p.~526).  Since
resonances are broader for short-lived waves, the quantitative gas
response may depend weakly on the duration of the wave, but any
attempt to use this as a discriminant between theories would require
an accurate independent estimate of the spiral amplitude.

\cite{GGF95} and \cite{FRZ10} estimate the gravity torque from the
spiral pattern.  If this measurement determines the rate of angular
momentum transport, then one can use it to estimate the time scale for
significant changes to the initial angular momentum distribution.  If
the spiral is long-lived, the gravity torque must be corrected for the
advective term (\S\ref{evolv}), which can substantially lengthen the
time-scale.  Advection is far less important for transient spirals.
While it is of great interest that the angular momentum evolution time
scale could be less than a Hubble time were spirals to be transient,
this measurement again does not discriminate between short- and
long-lived spiral patterns.

The ``density wave'' theory of spiral modes (\S\ref{theory}.3)
requires the outer disc to be dynamically cool, in particular Toomre's
axisymmetric stability parameter should be in the range $1 \la Q \la
1.2$.  In principle, a direct measurement of a larger value of $Q$ would
cause difficulties for this theory.  A number of attempts have been
made to estimate $Q$ in real discs \citep[\eg][]{Korm84,Bott93,HC09},
but uncertainties are large because the measurement of the low
velocity dispersion in a disc is challenging and the $Q$ value depends
on the disc surface density, which is not tightly constrained.  It
should also be noted that the raw $Q$-value estimated from the stars
only may not be the relevant quantity in realistic discs, where the
gas component and perhaps also the young stellar population can have a
disproportionate effect on the overall responsiveness \citep{Rafi01}.

\cite{DP10} suggest that the downstream age distribution of star
clusters can be used to determine the origin of a spiral pattern.
However, their method really is a test only of whether the gas that
formed the star clusters streamed through the spiral arm at a speed
that differs significantly from the local circular speed --
\ie\ merely whether a coherent density wave is present -- and does not
discriminate between mechanisms for its origin or test for its
lifetime.

\cite{Meid08} generalize the method of \cite{TW84} to discs that may
have multiple pattern speeds.  They \citep{MRM09} apply it to CO data
on a number of galaxies, finding evidence for multiple patterns in
several cases.  Other methods to estimate the pattern speed of spiral
arms often make the (implicit) assumption that there is a single
pattern; for example, the kinematics of stars formed in spiral arms
\citep{Bash79}, the change in radial flow direction across corotation
\citep[\eg][]{EWP98}, or the change in the pattern of residual
velocities \citep{Canz93}, or brute force modelling
\citep[\eg][]{RSL08}, \etc\ \ Once again, however, the results tell us
little about the lifetimes of the patterns.

In summary, I concur with BT08 that it is frustratingly difficult to
determine the lifetimes of spiral patterns, or the mechanism that gave
rise to them, from observations of external galaxies.

However, evidence from the Milky Way seems to favour the transient
spiral picture.  The in-plane velocity components of stars in the
solar neighbourhood as revealed by the {\small HIPPARCOS} mission
\citep{Dehn98,Nord04} do not have the smooth double Gaussian
anticipated by K. Schwarzschild (BT08, p.~321), but instead are
characterized by a number of separate streams, with essentially no
underlying smooth component \citep{BHR09}.  The features are too
substantial to have simply arisen from groups of stars that were born
with similar kinematics \citep[\eg][]{Egge96}, as confirmed in
detailed studies \citep{Fama07,Bens07,BH10}, and it is clear that the
entire \DF\ has been sculptured by dynamical processes.

Individual features in this distribution have been attributed to
resonances with the bar \citep{Rabo98,Dehn00,Fux01}, or with a spiral
pattern \citep{QM05,Sell10b}, while other models include both bars and
spirals \citep{Quil03,Chak07,Anto09}.  The overall appearance of the
velocity distribution was successfully reproduced by \cite{dSWT} who
invoked a succession of short-lived spiral transients.

\section{Spiral Structure Theory}
\label{theory}
Three distinct mechanisms have been proposed to account for
self-excited spiral patterns in disc galaxies.  \cite{BL96} propose
that spiral features are manifestations of quasi-steady global modes
of the underlying disc. \cite{GLB65} and \cite{Toom90} argue that,
aside from tidal- and bar-driven cases, most spirals are short-lived
transient features that are essentially collective {\it responses\/}
to internal density fluctuations within the disc.  \cite{Sell00} also
supposes that spirals are short-lived, but suggests they result from a
recurrent cycle of vigorous large-scale modes.  Here I give brief
summaries of each in turn; \cite{Sell10a} gives a more detailed
review.

\subsection{Swing Amplified Noise}
Swing amplification was first described by \cite{GLB65} and
\cite{JT66}.  \cite{Toom81} highlighted its importance, which he
illustrated with the now classic figure named ``dust-to-ashes''
reproduced in BT08 (their Fig.~6.19).  It shows the dramatic transient
trailing spiral that results from a small input leading disturbance.
\cite{JT66} showed that swing-amplification causes the collective
response of a disc to a co-orbiting mass clump to be a substantial
spiral ``wake.''

\cite{GLB65} and \cite{Toom90} suggested that a large part of the
spiral activity observed in disc galaxies is the collective response
of the disc to clumps in the density distribution.  As a spiral wake
is the collective response of a disc to an individual co-orbiting
perturber, multiple perturbers will create multiple responses that all
orbit at different rates.  The behaviour of this polarized disc reveals
a changing pattern of trailing spirals, which can equivalently be
regarded as swing-amplified noise.  \cite{TK91} show that amplified
noise arising from the massive disc particles themselves can be
understood in the shearing sheet, where the resulting spiral
amplitudes are linearly proportional to the input level of shot noise.

This could be a mechanism for ``a swirling hotch-potch of pieces of
spiral arms'' \citep{GLB65} in very gas rich discs, where a high rate
of dissipation may be able to maintain the responsiveness of the disc
\citep{Toom90} while the clumpiness of the gas distribution may make
the seed noise amplitude unusually high.  However, the spiral
structure should be chaotic, with little in the way of clear symmetry
expected.  Also, it seems likely that spiral amplitudes
\citep[\eg][]{ZCR09} are too large to be accounted for by this
mechanism in most galaxies.

\subsection{Recurrent Cycle of Groove Modes}
\cite{SK91} showed that discs are destabilized by a deficiency of
stars over a narrow range of angular momentum; in a disc without
random motion, such a deficiency would be a ``groove.''  The groove
itself is unstable, and the instability becomes a global mode through
the vigorous supporting response of the surrounding disc, producing a
large-scale spiral instability.  \cite{SK91} showed that the linear
instability, which is driven from corotation, also develops in discs
with random motion, while \cite{SB02} showed that the amplitude of the
mode was limited by the onset of horseshoe orbits.

\cite{SK91}, and earlier \cite{LH78}, found that almost any narrow
feature in the angular momentum density is destabilizing.  Thus the
common starting assumption of spiral structure studies, that the
underlying disc is featureless and smooth, may throw the spiral baby
out with the bathwater.

\cite{SL89} showed, in simulations of a low mass disc, that the
particles were scattered at the Lindblad resonances as each coherent
wave decayed, thereby leaving behind an altered distribution function
that created the conditions for a new instability.  \cite{Sell00}
described in more detail that each spiral pattern seems to be a
strongly unstable true mode of the disc in which it grows; it lasts
for only a few rotations, changing the disc properties as it decays so
as to provoke new instabilities of the altered disc having different
pattern speeds.  The discovery of resonance scattering among the stars
of the solar neighbourhood \citep{Sell10b} suggests that this
mechanism may occur in real galaxies.  I am actively engaged in
addressing the many details of this picture that remain unclear.

\subsection{Long-lived Global Modes}
Simple models of disc galaxies support many global linear instabilities
\citep[\eg][]{Kaln78,Jala07}.  The bar-forming mode is generally the
fastest growing, but has almost no spirality.  These studies are
therefore important to understand stability, but do not appear
promising for spiral generation.

The ``density wave'' theory for spiral modes, described in detail by
\cite{BL96}, invokes a more specific galaxy model with a cool outer
disc and hot inner disc.  The local stability parameter, $Q = \sigma_R
/ \sigma_{R,\rm crit}$ \citep{Toom64}, is postulated to be $Q \ga 1$
in the outer disc and to rise steeply to $Q > 2$ near the centre.
Under these specific conditions, Bertin \& Lin find slowly evolving
spiral modes that grow by their {\small WASER} \citep{Mark77}
mechanism.  They invoke shocks in the gas to limit the amplitude of
the slowly growing mode, leading to a quasi-steady global spiral
pattern.  \cite{Lowe94} present a model of this kind to account for
the spiral structure of M81.

The principal objection to their picture is that it is likely that an
outer disc with such a low $Q$ will support other, more vigorous,
collective responses that will quickly alter the background state by
heating the outer disc, as I now show.

\subsection{A Direct Test}
Here I report simulations designed to test directly whether a galaxy
model of the kind invoked by \cite[][hereafter BLLT]{Bert89} can in
fact survive to support the slowly-growing global mode they predict
should dominate.

\begin{figure}
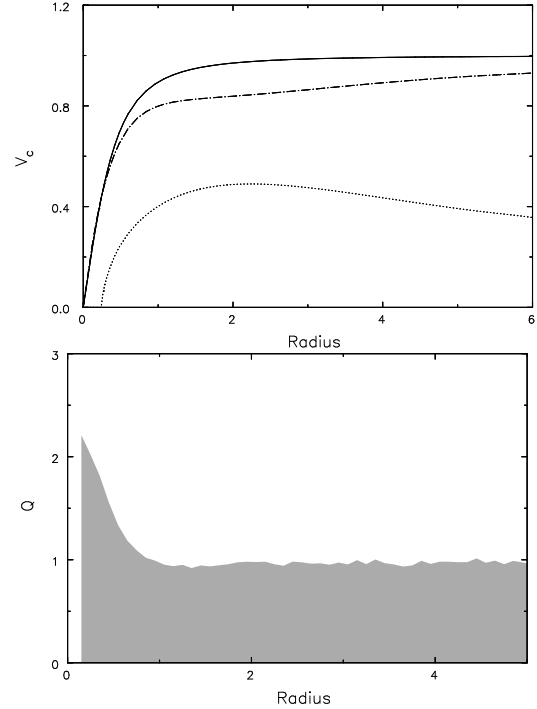

\begin{center}
\includegraphics[width=.82\hsize,clip]{rotc.ps}
\includegraphics[width=.8\hsize,clip]{qploti.ps}
\end{center}
\caption{Top panel shows the rotation curve of the model discussed in
  this section in units given in the text.  The solid line is the full
  circular speed, the dotted line shows the contribution from the
  modified exponential disc, and the dot-dashed line that from the
  halo.  The lower panel shows the initial radial variation of $Q$.}
\label{model}
\end{figure}

The model I adopt here is from the A$_1$ survey of BLLT, which is a
minor variant of that proposed by \cite{FE80}.  The surface density of
the disc is
\begin{equation}
\Sigma(R) = (1+\Delta){M_d \over R_d^2} e^{R/R_d}f(2R/R_d),
\end{equation}
with
\begin{equation}
f(y) = \cases{ 1 - (1+4y)(1-y)^4(1 - e^{-R/R_d}/6) & $y<1$ \cr
  1 & otherwise. \cr}
\end{equation}
Here $R_d$ and $M_d$ are respectively the scale length and mass, when
$\Delta=0$, of the unmodified exponential disc.  The $f(y)$ factor
creates a central dip in the surface density of approximate radius
$R_d/2$.  The rotation curve of their model has the form
\begin{equation}
V(r) = V_0 {x \over (1 + x^2)^{1/2}} \quad\hbox{with}\quad x = 2R/R_d,
\end{equation}
that is independent of the disc contribution.  Because mass removed by
the central cutout is effectively replaced by rigid matter, the model
can be thought of as including a small central bulge as well as a
pseudo-isothermal halo.  Here I adopt units such that $G=M_d=R_d=1$,
and also $V_0=1$; the orbit period at $R=2R_d$ is $\sim13$ time units.

BLLT also specify the radial variation of Toomre's local axisymmetric
stability parameter $Q(R) = Q_{\rm OD}(1 + 1.5e^{2R/R_d})$;
\ie\ $Q=2.5Q_{\rm OD}$ at the centre and decreases rapidly to
$Q=Q_{\rm OD}$ over most of the outer disc.  Because BLLT do not
supply stellar dynamical distribution functions, I set the initial
velocities to create a rotationally supported disc having the desired
$Q$ profile, using the Jeans equations in the epicyclic approximation
to determine the azimuthal dispersion and asymmetric drift (BT08,
eq.~4.228).

I have chosen the case with $Q_{\rm OD}=1$ and $\Delta = -0.35$.  The
upper panel of Fig.~\ref{model} shows the rotation curve, which
indicates that this model has a strongly sub-maximal disc
\citep{Sack97}, \ie\ the disc contributes merely $\sim25\%$ of the
central attraction at $R=2R_d$.  The lower panel shows the initial
$Q$-profile.  Because the disc is so light, and the disc surface
density dips towards the centre, the initial random velocities are
small enough that the epicycle approximation is adequate and the Jeans
equations lead to a quite respectable equilibrium model.

BLLT predict that this model has the slowly-growing, tightly-wrapped,
bi-symmetric spiral mode illustrated second from left in the bottom
row of their Fig.~3, but they do not give the frequency of the mode.

\begin{figure}
\begin{center}
\includegraphics[width=.9\hsize,clip]{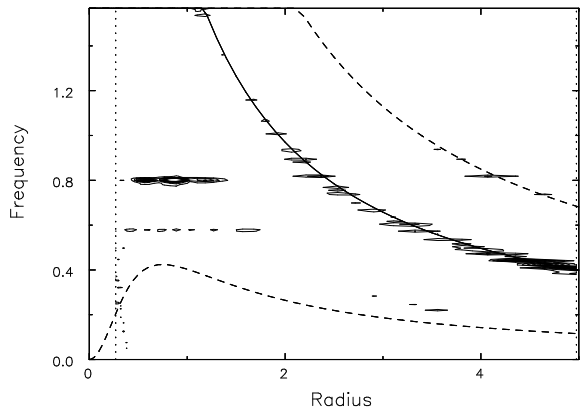}
\end{center}
\caption{Contours of power as a function of radius and frequency
  computed from the relative bi-symmetric density variations
  ($\Sigma_2/\Sigma_0$) on the rings of the simulation grid.  Peaks
  indicate coherent features with pattern speed $m\Omega_p$ that are
  present for a significant fraction of the simulation.  The solid
  curve shows the circular angular frequency $m\Omega$, and the dashed
  curves $m\Omega \pm \kappa$.  The simulation was run for 500 time
  units, or $38.5$ disc rotations at $R=2$.}
\label{spctB}
\end{figure}

Here I report the results of two series of simulations with this disc
model.  In the first series, I restrict disturbance forces to $m=2$
only, as do BLLT in their mode calculations.  In the second, I allow
sectoral harmonics $2 \leq m \leq 8$ to contribute.  In neither case
do I include disturbance forces from $m=0$ \& 1.  Excluding $m=1$
avoids unbalanced disturbance forces that could arise from a lop-sided
mass distribution in a rigid halo; there are no external perturbations
to excite such waves, which are also most unlikely to be self-excited
in this low mass disc (see below).  The unchanging axisymmetric
central attraction maintains the rotation curve shown in
Fig.~\ref{model}.

The simulations employ the grid code that has been shown to reproduce
global modes predicted by linear theory for a number of stellar
dynamical models \citep{SA86,Sell89,ES95,SE01}.  The particles move
over a 2D polar grid having $65 \times 96$ mesh points and their
motion is integrated with a time step of 0.05.  Direct tests reveal
that results are insensitive to these numerical parameters.  The
particles interact with forces derived from a Plummer softening kernel
with scale $0.05R_d$.

With $m=2$ only disturbance forces, a simulation with $N=200\,000$
particles barely evolved.  No non-axisymmetric features were visible
in the first 40 disc rotations and the $Q$-profile remained unchanged.
There is a hint in the power spectrum shown in Fig.~\ref{spctB} of a
coherent wave in the inner disc rotating at the angular rate
$m\Omega_p \simeq 0.8$, but it is barely stronger than the noise peaks
along the $m\Omega$ curve.  Without the predicted eigenfrequency, I am
unable to confirm whether it corresponds to the {\small WASER} mode
predicted by BLLT.  Nevertheless, it is clear that the model does not
possess any rapidly-growing bi-symmetric instabilities.

\begin{figure}
\includegraphics[width=\hsize,clip]{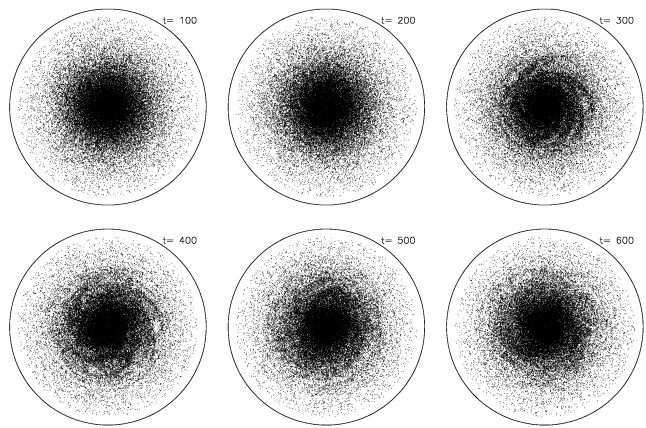}
\caption{The time evolution of the simulation in which disturbance
  forces included sectoral harmonics $m>2$.  Only one particle in 400
  is plotted in order to avoid saturating the images.  Multi-arm
  patterns appear soon after the start.  The grid edge (outer circle)
  has radius $R=6.5$ and one rotation at $R=2$ is 13 time units.}
\label{pntsB}
\end{figure}

\begin{figure}
\begin{center}
\includegraphics[width=.8\hsize,clip]{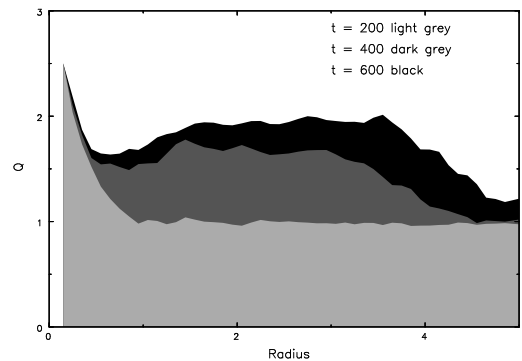}
\end{center}
\caption{The radial variation of $Q$ at three times in a simulation
  with $20$M particles and sectoral harmonics $2\leq m\leq8$ all
  active.  The initial $Q$-profile is that shown in Fig.~\ref{model}.}
\label{qplotn}
\end{figure}

The outcome is completely different when disturbance forces from
higher sectoral harmonics are included, as shown in Fig.~\ref{pntsB}.
Irrespective of the number of particles, or whether the particles are
positioned carefully on rings (a quiet start) or placed at random, the
model quickly develops multi-arm spiral patterns that heat the disc,
as shown in Fig.~\ref{qplotn}.

The reason for this different behaviour when higher sectoral harmonics
are included is clearly related to swing-amplification, which is
strongly dependent on the value of $X \equiv Rk_{\rm crit}/m$
\citep[][BT08]{JT66,Toom81}, where $k_{\rm crit} \equiv \kappa^2 / (2\pi
G\Sigma)$, the smallest wavenumber of axisymmetric Jeans instabilities
\citep{Toom64}.  In the present model, selected from BLLT, this
parameter rises steadily with radius from $X \simeq 8/m$ at $R=1$ to
$X \simeq 20/m$ at $R=3$, continuing to higher values farther out.
Since the swing-amplifier is vigorous only when $1 \la X \la 2.5$ in a
flat rotation curve, as here, it is pretty much dead at all radii for
$m=2$ waves, but vigorous responses are expected for $m \geq 4$, as we
observe.

\begin{figure}
\begin{center}
\includegraphics[width=.8\hsize,clip]{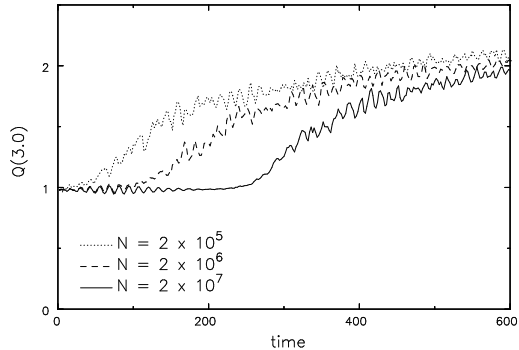}
\end{center}
\caption{The time evolution of $Q$ at $R=3R_d$ in three simulations
  with differing numbers of particles.  Sectoral harmonics $2\leq
  m\leq8$ were active in all three cases.}
\label{Qtplot}
\end{figure}

Fig.~\ref{Qtplot} shows the time evolution of $Q$ at $R=3R_d$ in three
simulations with differing numbers of particles showing that heating
of the disc, which coincides with the occurrence of visible spiral
patterns, is increasingly delayed as larger numbers of particles are
employed.  The principal effect of increasing $N$ is to reduce the
amplitude of initial density fluctuations.  The swing-amplifier
quickly polarizes the disc, creating trailing spiral responses having
an amplitude proportional to the input noise signal \citep{TK91}.  The
larger the particle number the smaller the initial amplitude, and it
is clear from Fig.~\ref{Qtplot} that the swing-amplified noise has too
small an amplitude to cause heating at first when $N=20$M.  We do not
expect global instabilities in a smooth disc for $m>2$ because
small-amplitude disturbances at most reasonable pattern speeds will be
damped at an \ILR\ \citep{Mark74}.  Thus the later increase in
the spiral amplitude must be a non-linear effect, perhaps related to
the recurrent instability cycle reported by \cite{SL89}.  Whatever the
cause, the similar heating rate and final $Q$ value is consistent with
spiral amplitudes that are independent of $N$ at later times.  The
origin of multi-arm spiral waves will be followed up in future work.

Fig.~\ref{Qtplot} shows that even with $N=20$M, heating begins after
just 15 disc rotations.  This should be contrasted with the case where
forces were restricted to $m=2$ where no visible bi-symmetric features
appeared in 40 rotations even with the ten times larger seed amplitude
implied by 100 times fewer particles.  It should be further noted that
real spiral galaxy discs have large star clusters and giant molecular
clouds that imply much larger seed density fluctuations than arise
from $N=20$M randomly-placed, equal-mass particles.

\subsection{Discussion}
While these simulations explicitly test just one of the many models
presented in BLLT, all the cases that they regard of most interest
have strongly sub-maximal discs in order that swing amplification is
ineffective for $m=2$; this aspect is essential so that the model
possess only slowly-growing, tightly-wrapped, bi-symmetric spiral
modes.  However, the vigour of the swing-amplifier for $m>2$ makes it
inevitable that every one of their galaxy models with a low-mass disk
will be subject to stronger activity due to disturbance forces with
$m>2$, which will quickly heat the outer disc and destroy the
conditions they require, as just demonstrated.  Thus the ``basic
state'' invoked by BLLT for the modes they favour to account for
bi-symmetric spiral patterns could not survive in real discs that
permit disturbances of all sectoral harmonics.

As this conclusion depends largely on the results from simulations,
one must worry whether they can be trusted.  The principal source of
concern is that the origin of the multi-arm patterns that heat the
disc remains obscure.  Simulations of this kind over many years
\citep[\eg][]{SC84,Fuji10} have manifested recurrent multi-arm spiral
patterns, and the behaviour has not changed as numerical quality has
risen and the codes have passed many tests.  However, the possibility
that the behaviour could result from some artefact in the simulations
cannot be excluded altogether until a satisfactory explanation is
provided.  Note that to doubt the result on these grounds calls into
question all simulations of isolated discs over the past 40 years, as
well as those that model the formation of disc galaxies.

Other possible criticisms, such as the simulations could be unreliable
because of gravity softening, particle noise, or the restriction to
2D, are more readily rebutted.  First, the very same code has been
shown to reproduce modes predicted from linear stability analyses of
other models, as cited above.  Also, Plummer softening in simulations
with particles confined to a plane provides a reasonable allowance for
finite disc thickness and anyway the same behaviour persists in
simulations of discs with finite thickness \citep{Rosk08,Fuji10}.
Further, since the only variation in the behaviour as $N$ is increased
100-fold is an increasing delay due to a decreasing seed amplitude,
with no other qualitative differences, an argument that simulations
cannot be trusted because $N$ is too small is somewhat threadbare.

One might also worry that the stellar dynamical realization I have
created differs from the model that BLLT analyzed in the
hydrodynamic approximation -- essentially I have replaced pressure
in their hydrodynamic calculations with velocity spreads; a more
direct test would require a prediction of a slowly-growing spiral in a
stellar-dynamical model.  However, it is hard to see why this minor
difference should matter, especially in this case where the dispersion
is a small fraction of the orbit speed almost everywhere; the
exception is at the centre where $Q$ is so large that the disc is (by
design) dynamically inert.

The simulations here have not, of course, made any allowance for the
influence of the gas component, which can offset the heating effects
from transient spiral patterns to some extent.  However, \cite{SC84}
found that simulations that included a very crude form of cooling
still settled to $1.5 \la Q \la 2$.

\section{Simulations}
\label{sims}
Short-lived recurrent spiral patterns have developed spontaneously in
simulations of isolated galaxies, from the first studies by
\cite{MPQ70} and \cite{HB74}, right through to modern simulations that
include more sophisticated physical processes
\citep[\eg][]{Rosk08,ATM10}.

Claims of long-lived spiral waves \citep[\eg][]{TEDS} have mostly been
based on simulations of short duration.  For example, \cite{ET93}
presented a simulation that displayed spiral patterns for $\sim 10$
rotations, but the existence of some underlying long-lived wave is
unclear because the pattern changed from snapshot to snapshot.  Other
claims are equally doubtful, as I show next.

\subsection{Direct Tests}
As \citet[][hereafter DT94]{DT94} and \citet[][hereafter Z96]{Zhan96}
have presented evidence for long-lived spirals in the same model, I
have chosen to try to reproduce their results here.  I first summarize
the model they employed and then report my own analysis of the similar
results I obtain when I reproduce their simulations.

\begin{figure}
\begin{center}
\includegraphics[width=.8\hsize,clip]{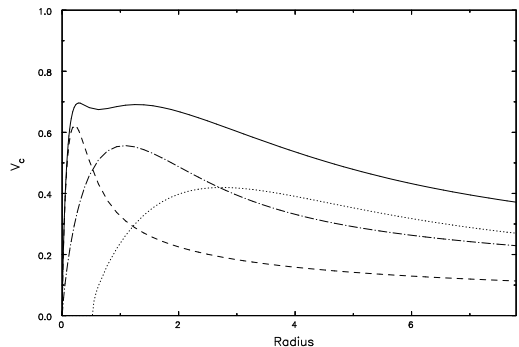}
\end{center}
\caption{The rotation curve of the initial model adopted by DT94 and
  Z96 in units given in the text.  The solid line is the full circular
  speed, the dotted line is the speed due to the modified exponential
  disc, and the dashed and dot-dashed lines show the speed due the
  bulge and halo respectively.}
\label{rotcZ}
\end{figure}

DT94 adopted the disc surface density distribution \citep{RK80}
\begin{equation}
\Sigma(R) = {2M_d \over 3\pi R_d^2} \left[ e^{-R/R_d} - e^{-2R/R_d} \right].
\end{equation}
Here $R_d$ is the scale length of the outer exponential disc and $M_d$
is the disc mass.  DT94 and Z96 chose $M_d = 0.5M_t$, where $M_t$ is
the total mass of the model, and employed two additional mass
components to represent a central bulge and a halo, both of which
exert the central attraction in the mid-plane of a razor-thin simple
exponential disc.  The masses and scale lengths were respectively
$0.1M_t$, $0.1R_d$ for the bulge and $0.4M_t$, $0.5R_d$ for the halo.
The rotation curve of this model is shown in Fig.~\ref{rotcZ}, which
compares well with that shown in Fig.~1 of DT94.  These authors set
the initial velocities in the disc such that $Q=1$ at all radii.

Here I recreate this model, and compute its evolution using
essentially the same 2D polar grid code, but with a larger number of
particles.  The disc has $N=2$M particles that move over a grid having
$100 \times 128$ mesh points.  As DT94 and Z96, I use a Plummer
softening law with a length scale $0.15R_d$ to compute forces between
particles.  However, I adopt a more physically motivated set of units
for which $G = M_t = R_d = 1$.  For comparison with the previous
results, it should be noted that $R_d = 10$ in their units, and one
rotation period at $R=2R_d$, which takes $2\pi R/V_c = 18.5$ of my
time units (925 time steps), is 314 time steps in DT94 and 628 time
steps in Z96.

\begin{figure}
\includegraphics[width=\hsize,clip]{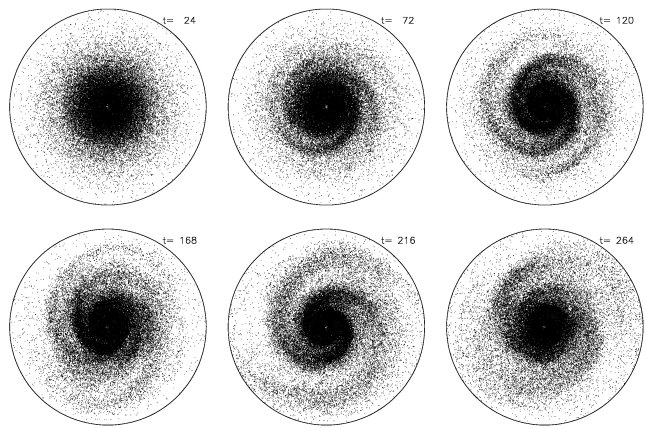}
\caption{Evolution of the model run to reproduce that reported by DT94
  and Z96.  The six snapshots, which include one particle in 40,
  should be compared with those shown in Fig.~2 of both papers.  The
  circle is drawn at $R=8R_d$, although the grid extends to
  $12.7R_d$.}
\label{pntsZ}
\end{figure}

Fig.~\ref{pntsZ} shows the evolution computed here, which should be
compared with that shown in Fig.~2 of Z96.  Since the ten times larger
number of particles used here lowers the seed amplitude, a little more
evolution is needed for the spiral to grow.  To make the closest
possible comparison, I therefore show snapshots that are spaced at the
same time interval, but are shifted later by a little over one disc
rotation from the start.  The overall appearance is quite similar; a
strong $m=2$ spiral is developing by time 72 and is perhaps more
persistent than that in Zhang's calculation, where the spiral has
faded more by the last two times.

\begin{figure}
\begin{center}
\includegraphics[width=.9\hsize,clip]{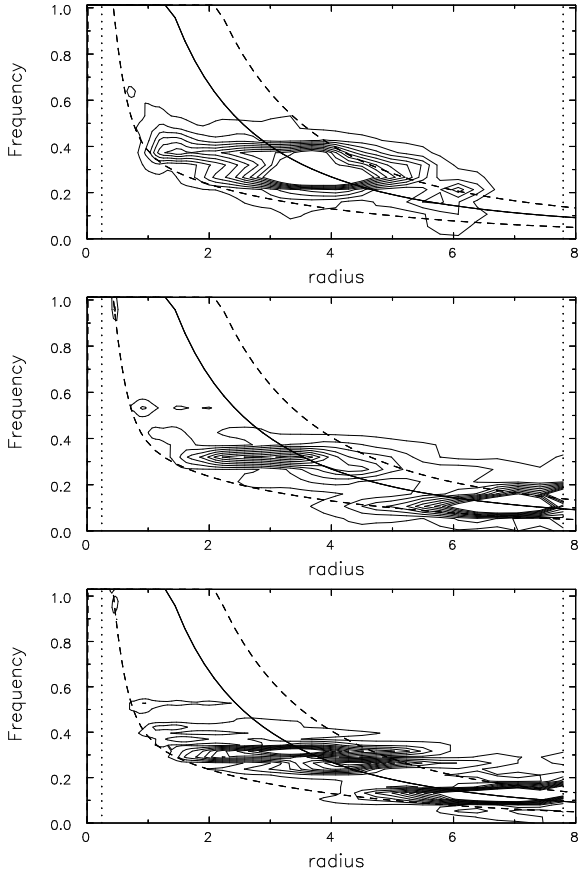}
\end{center}
\caption{Contours of power as a function of radius and frequency
  computed from the relative bi-symmetric density variations
  ($\Sigma_2/\Sigma_0$) on the rings of the simulation grid.  Peaks
  indicate coherent features with pattern speed $m\Omega_p$ that last
  for a significant fraction of the simulation.  The solid curve
  shows the circular angular frequency $m\Omega$, and the dashed curves
  $m\Omega \pm \kappa$.  The top panel is computed from the first half
  of the simulation, the middle panel from the second half, while the
  third panel shows the spectrum from both halves combined.}
\label{spctZ}
\end{figure}

Since DT94 and Z96 claim that the $m=2$ features are a long-lived
spiral, I examine them more closely here.  Fig.~\ref{spctZ} shows that
in my simulation they appear to be the super-position of several waves
having differing pattern speeds.  This figure should be compared with
Fig.~5 of DT94, which presents a similar analysis for their model.

Although I employed 40 times the number of particles used by DT94, the
power spectrum is still quite noisy.  However, the lowest panel has
three or more horizontal ridges that are caused by coherent waves
extending roughly from the \ILR\ to a little outside the \OLR\ in each
case.  The lowest frequency peak, which is the farthest out in the
disc, has the largest relative amplitude, which is simply a reflection
of the fact that the disc surface density decreases outward.  Forming
separate power spectra on the first and second halves of the evolution
(top two panels of Fig.~\ref{spctZ}) reveals that the separate
patterns reach peak amplitude in sequential order, with the fastest
rotator ($m\Omega_p=0.42$) developing and decaying first -- there is
no significant power at that frequency in the second half of the run.
At least two waves co-exist at significant amplitude for most of the
evolution.

A least-squares fit to these data \citep{SA86}, as well as to an
expansion of the particle distribution in logarithmic spirals, finds
at least four coherent waves with $m\Omega_p \simeq 0.42$, 0.30, 0.25
\& 0.12, in rough agreement with the locations of the peaks in the
lowest panel of Fig.~\ref{spctZ}.  These values need to be multiplied
by $18.5/314 \approx 0.06$ to convert to the frequency unit used by
DT94, and these authors plot pattern speed $\Omega_p$ and not
$m\Omega_p$ that I measure; thus the pattern speeds I observe are
0.013, 0.009, 0.007 \& 0.004 in the units shown in Fig.~5 of DT94,
where three waves can also be discerned.  The fastest and most slowly
rotating waves in DT94 have similar frequencies to those I observe,
but neither intermediate wave corresponds to a feature in their plots.
Exact correspondence should not be expected as all these waves are
seeded by particle noise.

\begin{figure}
\begin{center}
\includegraphics[width=.9\hsize,clip]{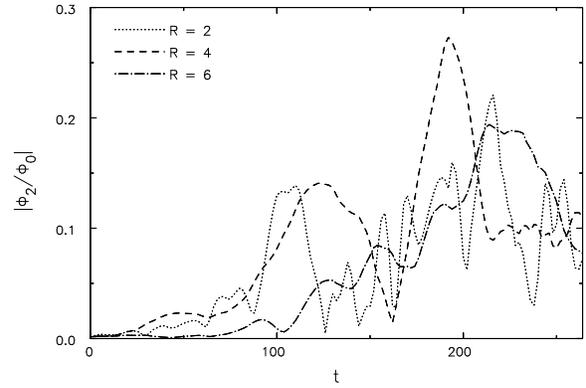}
\end{center}
\caption{The time variation of the relative amplitude of the
  bi-symmetric {\it potential\/} at three different radii.}
\label{getpot}
\end{figure}

Fig.~\ref{getpot} confirms that the amplitude of bi-symmetric
disturbances rises in the inner disc (dotted line) earlier than in the
outer disc (dot-dash line).  Furthermore, the potential variations
fluctuate in amplitude, especially in the inner disc, in a manner
characteristic of beats between waves rotating at different pattern
speeds, as to be expected from the power spectra shown in
Fig.~\ref{spctZ}.  Were the evolution dominated by a single $m=2$
spiral mode, the temporal evolution in this Figure would be smooth and
vary synchronously at each radius.

Thus ``the bisymmetric spiral'' in this simulation is not a single
long-lived pattern, but the superposition of three, or more, waves
that each grow and decay.  \cite{Zhan98} argues that this model has a
tendency to form a bar, which I have also noticed, and she therefore
presents another simulation in which the disc mass is reduced to
$0.4M_t$ and the halo mass increased to $0.5M_t$ in order to weaken
the tendency to form a bar.  As she again claims that this new model
supports a long-lived spiral, I have also tried to reproduce this
slightly different simulation.  As above, I find that the strong, and
visually impressive, bi-symmetric disturbance is the superposition of
more than two patterns that each do not last long.  In this case the
first wave to appear has frequency $2\Omega_p \approx 0.29$, but later
the two dominant waves have $2\Omega_p \approx 0.36$ \& 0.19, with
also some significant amplitude at $m=3$.

\section{Indirect Evidence}
\label{indirect}
Here I summarize three further arguments that bear indirectly on the
question of spiral lifetimes.  All three are related to the effect of
spiral activity on the orbits of stars in the disc.

\subsection{Importance of Gas}
Transient spiral activity scatters disc stars away from circular
orbits \citep{BW67,CS85,BL88}.  As random motion of the disc stars
rises, the collective behaviour that produces spiral patterns is
weakened.  Thus, in the absence of cooling, spiral activity is
self-limiting -- it heats the disc to a certain level at which the
disc can no longer support spiral patterns.

\cite{SC84} showed this explicitly in their simulations, and went on
to show that a moderate amount of cooling in the form of fresh stars
added to the disc on circular orbits enabled spiral activity to
continue ``indefinitely.''  Subsequent work on isolated discs
\citep{CF85,Toom90,Rosk08}, as well as modern galaxy formation
simulations \cite[][and many others]{ATM10}, has confirmed this result
in simulations with ever greater realism.  Physically, the random
motions gained by gas clouds are dissipated in collisions so that they
keep moving on near circular orbits.  Stars that form from the clouds
therefore have similar motions, and a continuous supply of fresh stars
on near-circular orbits maintains a responsive stellar distribution
that allows spiral activity to continue.  Quantitatively, fresh stars
added to a disc at the rate of a few every year is sufficient for this
purpose.

Thus the transient spiral picture offers a natural explanation for the
absence of spiral patterns in S0 disc galaxies that have little or no
gas and the long-noted correspondence that disc galaxies with
significant gas components also manifest spiral patterns.  Note that
this appealing argument again does not uniquely favour short-lived
waves, since \cite{BL96} also invoke a gas component both to limit
the amplitude of their mildly growing modes and to maintain the
dynamically cool outer disc.

\subsection{Scattering of Stars}
It has been clear for some time that the velocity dispersion of disc
stars in the solar neighbourhood rises with age \citep{Wiel77,Nord04}
and also, for main sequence stars, with colour \citep{AB09}, which is a
surrogate for mean age.  The highest velocities cannot be produced by
cloud scattering \citep{Lace91,HF02} and some other accelerating
agent, such as transient spirals, seems to be required.  Long-lived
spirals, which do not heat the disc nearly as effectively, could not
achieve the requisite high velocities.

\subsection{Radial Mixing}
Studies of the metallicities and ages of nearby stars
\citep{Edva93,Nord04,Reid07,Holm07} find that older stars tend to have
lower metallicities on average.  As it is difficult to estimate the
ages of individual stars, the precise form of the relation is still
disputed.  However, there seems to be general agreement that there is
a spread of metallicities at each age, which is also supported by
other studies \citep{CHW03,Hayw08,SH10}.  A metallicity spread amongst
coeval stars is inconsistent with a simple chemical evolution model in
which the metallicity of the disc rises monotonically in each annular
bin, without mixing in the radial direction.  Again \cite{SB02} and
Ro\u skar \etal\ (2008ab) showed that when the disc supports recurrent
transient spirals, the needed radial mixing arises naturally through
angular momentum changes at corotation that do not heat the disc.
\cite{SB09} developed the first chemical evolution model for the Milky
Way disc to include this radial churning.

The transient nature of the patterns is an essential aspect to produce
efficient churning.  The horse-shoe orbits near co-rotation of a
large-amplitude spiral cause a single change in the angular momentum
if the spiral grows and decays in less than half the (long) orbit
period {\it in the rotating frame}.\footnote{In fact, \cite{SB02}
  argue that the spiral mode grows until it reaches the amplitude at
  which the periods of horse-shoe orbits become short enough for this
  to happen for many stars, which {\it causes\/} the disturbance to
  disperse.}  Were the spiral to persist for longer than this, then
the exchanges a star experiences at one arm would be undone when the
star encounters the next arm, and the star would merely take periodic,
equal inward and outward steps in radius leading to no lasting change.
\cite{Minc10} report substantial mixing due to combined influence of a
bar and spirals, but their simulations were of short duration and it
is likely the effect is large only as the bar forms.  Only recurrent
transient patterns, with corotation radii spread at random locations,
can cause the home radii of stars to ``diffuse'' across the disc
throughout its life.

\subsection{Discussion}
Tremaine (private communication) pointed out that transient spirals
could co-exist with long-lived patterns.  This possibility could be
consistent, for example, with the change in the appearance of spiral
patterns between the blue and near-IR passbands
\citep[\eg][]{BW91,Bloc94}, although these data contain no information
about the lifetimes of features of any colour.  The co-existence of
short-lived spirals seems inconsistent with the specific mode theory
for long-lived patterns proposed by \cite{BL96}, since transient waves
would heat the outer disc to an extent that must render their proposed
mechanism unworkable, as shown in \S\ref{theory}.4.  However, some
other possible (as yet unknown) mechanism for long-lived spirals might
remain viable.

An observational test for the coexistence of both short- and
long-lived patterns would be a daunting challenge.  For a
well-constructed sample of galaxies, one would have to determine the
mean amplitude needed to produce the desired heating and churning by
the short-lived patterns, and then show that the mean observed spiral
amplitudes, estimated somehow from the passband dependent photometry,
would or would not allow an additional significant long-lived spiral
component.

\section{Conclusions}
The question of the lifetimes of spiral patterns is important because
galaxy discs evolve to a much greater extent if spirals are transient
than if they are quasi-steady (\S\ref{evolv}).  In this paper, I have
summarized the evidence that bears on the question; were any one piece
decisive, there would have been no need to write this paper.  However,
I find that the transient picture is favoured by the velocity
structure in the solar neighbourhood (\S\ref{observ}) and by the
behaviour of simulations (\S\S\ref{theory}.4 \& \ref{sims}).

Direct observational evidence from external galaxies (\S\ref{observ})
is frustratingly inconclusive, largely because our single snapshot
view can tell us little about the lifetimes of the patterns or, if
self-excited, the mechanism that caused them.  However, the overall
appearance of the velocity distribution of stars in the solar
neighbourhood is most naturally accounted for by a succession of
transient spirals, with individual features perhaps being attributable
to resonances.

Theoretical work (\S\ref{theory}) over many decades has pursued the
two mutually exclusive views.  In their efforts to gain the upper
hand, the rivalry has spurred great improvements in our understanding
of density wave mechanics.  However, neither side has been able to
land a knockout blow.

Simulations have long manifested short-lived transient features, and I
have presented two new results here.  The first, in \S\ref{theory}.4,
shows that galaxy models with cool, low-mass discs of the type
proposed by \cite{Bert89} as likely to manifest long-lived,
bi-symmetric spiral patterns will instead heat quickly as a result of
short-lived transient spiral patterns with $m>2$.  While the behaviour
in the simulations is not fully understood, swing-amplification gives
a clear reason for the preference for higher sectoral harmonic
disturbances in low-mass discs.  Naturally, my demonstration that the
most carefully worked-out theory for long-lived modes may not be
viable, does not exclude the possibility of long-lived modes
altogether.

I have also shown (\S\ref{sims}.1) that the most clearly presented
case of a long-lived spiral in a simulation is not, in fact, a single
spiral mode, but a sequence of a few strong patterns that each have
short lifetimes.  Thus I am unaware of a credible case of a long-lived
spiral in any simulation, and therefore simulation results provide
strong support for the transient hypothesis.  The principal reason
this is not decisive is that we still do not fully understand the
origin of spiral patterns in the simulations.

Indirect evidence (\S\ref{indirect}) suggests that three quite diverse
properties of galaxies are naturally explained by transient spiral
waves, but an underlying long-lived wave is a possibility that would
be very difficult to exclude observationally.

To hold that spirals are long-lived, one must also argue, (1) that
significant long-lived spirals underlie the transients that appear to
be needed to explain the velocity structure in the solar
neighbourhood, the high random motions of old disc stars and radial
mixing in discs, (2) that the long-lived pattern adjusts its rotation
rate slowly over time, so that fresh stars become the sources and
sinks of angular momentum to drive the pattern, and (3) that no
simulation has yet been able to capture the correct dynamical
behaviour of spiral patterns.

However, the entire question of spiral lifetimes will not be settled
until a theory of spiral pattern generation, which is supported by
observational data and consistent with simulation results, gains wide
acceptance.

\section*{Acknowledgments}
I thank the referee (Gene Byrd), Xiaolei Zhang, and especially
Giuseppe Bertin for comments on a draft of this paper that helped to
clarify some issues.  I also thank Scott Tremaine for encouragement
and for a number of perceptive remarks.

\label{lastpage}

\end{document}